\documentclass[aps,prb,twocolumn,showpacs]{revtex4-1}  

\usepackage{graphicx}  
\usepackage{dcolumn}   
\usepackage{bm}        
\usepackage{amssymb}   
\usepackage{mathrsfs}
 
\newcommand\R{\mathscr{R}}
\def\journal#1, #2, #3, #4#5#6#7{
   {{#1}~{\bf{#2}},          
    #3 (#4#5#6#7).}}%
\def\jourpar#1, #2, #3 (#4){\journal #1, #2, #3, #4 \par}
\newcommand\PRB{\journal Phys. Rev. B, }%
\newcommand\PRL{\journal Phys. Rev. Lett., }%
\newcommand\josaB{\journal J. Opt. Soc. Am. B, }%

\begin{document}

\title{Use of X-ray scattering functions in Kramers-Kronig analysis of reflectance}
\author{D.B. Tanner}
\affiliation{Department of Physics, University of Florida, Gainesville, Florida 32611, USA}
\date{\today}

\begin{abstract}
Kramers-Kronig analysis is commonly used to estimate the optical properties of new materials. The analysis typically uses data from far infrared through near ultraviolet (say, 40--40,000 cm$^{-1}$ or 5 mev--5 eV) and uses extrapolations outside the measured range. Most high-frequency extrapolations use a power law, 1/$\omega^n$, transitioning to $1/\omega^{4}$ at a considerably higher frequency and continuing this free-carrier extension to infinity. The mid-range power law is adjusted to match the slope of the data and to give pleasing curves, but the choice of power (usually between 0.5 and 3) is arbitrary. Instead of an arbitrary power law, it is is better to  use X-ray atomic scattering functions such as those presented by Henke and co-workers. These basically treat the solid as a linear combinations of its atomic constituents and, knowing the chemical formula and the density, allow the computation of dielectric function, reflectivity, and other optical functions. The ``Henke reflectivity'' can be used over photon energies of 10 eV--34 keV, after which a $1/\omega^{4}$ continuation is perfectly fine. The bridge between experimental data and the Henke reflectivity as well as two corrections that needed to be made to the latter are discussed.
\end{abstract}

\maketitle

\section{Introduction}

Kramers-Kronig\cite{Kronig,Kramers,jackson,LandauLifshitz,wooten,DresselGruner} analysis, of---for the most part---reflectance data, is often used to estimate the optical conductivity, dielectric function, sum rules, and other optical functions for new materials. 
Many reports of Kramers-Kronig analysis of reflectance have appeared, spanning more than  50 years,\cite{Jahod57pr} with studies of metals,\cite{Taft61pr,Ehrenreich62pr,Ehrenreich63pr,Irani71josa,Hagemann75,Winsemius76,Shiles80prb,Leveque83} pure and doped elemental solids,\cite{Phillip64pr,Taft65pr,Tongay10prb} organic conductors,\cite{Jacobsen74prl,Jacobsen81prl,Dressel96prl} charge-density-wave materials,\cite{Beal79jpc,Greene82prb,Challener84ssc} conducting polymers,\cite{Fincher79prb,Stafstrom87prl,Lee06nat} cuprate superconductors,\cite{Bonn87prl,Reedyk88prb,Kamaras90prl,cooper90prb1,gao93prb,cooper93prb,homes93prl,Basov94prb2,Basov95prl,Quijada99prb,Puchkov96prl} manganites,\cite{Kaplan96prl,Murakami98prl,Kim98prl} pnictides,\cite{Li08prl,Lucarelli10njp,Barisic10prb,Cheng12prb,Dai12prb,Moon13prl,Dai13epl} heavy-Fermions,\cite{Bonn88prl}
 multiferroics,\cite{Pimenov06prb,Xu08prl,Miller12prb,Miller13prb} topological insulators,\cite{LaForge10prb,Akrap12prb,Martin13prb} and many others. In addition, a number of methods papers have 
appeared.\cite{Roessler65bjap,Ahrenkiel71josa,Bardwell85jcp,Peiponen91prb,King02josab,Lucarini05,Kuzmenko05rsi,Crandles08ao} 

The experimenter typically has data from far infrared through near ultraviolet, covering, say, 5 mev--5 eV (40 to 40,000 cm$^{-1}$). This is a reasonably wide bandwidth, but the Kramers-Kronig integral extends from zero to infinity, so that extrapolations need to be made outside the measured range. The high frequency extrapolation is especially problematic and can cause significant distortions to the conductivity over the entire measured range, with consequences for sum rules as well. The approach used by most is to extend the reflectance with a  power law, $\R \sim 1/\omega^n$, transitioning to $\R \sim 1/\omega^{4}$ at a considerably higher frequency and continuing this free-carrier extension to infinity. The mid-range power law is adjusted to match the slope of the upper end of the data and to give pleasing curves, but the choice of power (something between 0.5 and 3) is arbitrary.

Other approaches have been put forward. One heroic method is to carry out ellipsometry on the sample over the high-energy part of the interesting spectral range and extract the (temperature-dependent if necessary) complex refractive index over that range. Then one can calculate an oscillator-model extrapolation that forces the Kramers-Kronig-derived refractive index to agree with ellipsometry over the range of overlap.\cite{Bozovic90prb,Wu93jp,Kamaras95jap,Puchkov96jp,Kuzmenko02ssc,Carbone06prb,Stricker14prl}
A second approach\cite{Kuzmenko05rsi} consists of fitting the spectrum with a sum of a very large number of narrow contributions to the dielectric function. The functions can be  Lorentz oscillators, triangles, or some other function for the imaginary part and the Kramers-Kronig-derived counterpart for the real part. The number of these functions is equal to or nearly equal to the number of data points, so that an excellent fit is easy to obtain; indeed some parameters need to be fixed. The model dielectric function then represents the properties of the material. No actual integral of the reflectance is computed. This approach is especially effective for the case of a thin film on a substrate\cite{Kuzmenko05rsi,Crandles08ao} or a complex device structure.\cite{Hwang11prb}

This paper describes an extrapolation method for conventional Kramers-Kronig analysis that uses X-ray atomic scattering functions developed by Henke and co-workers\cite{henke93,henkeweb} to generate the high-frequency reflectance of a material. The method basically treats the solid as a linear combination of its atomic constituents. Knowledge of the chemical formula, the density, and the scattering function enables the computation of the dielectric function, the reflectivity, and other optical functions. The ``Henke reflectivity'' is computed for photon energies of 10 eV--34 keV, after which a $1/\omega^{4}$ continuation is perfectly fine. This paper also discusses the bridge between experimental data and the Henke reflectivity as well as two corrections that needed to be made to the latter.

\subsection{The Kramers-Kronig relations}

The Kramers-Kronig relations\cite{Kronig,Kramers} are a consequence of our experience that observable effects are {\it causal\/}, {\it i.e.,\/} that the cause precedes the effect.  This notion seems sensible and it is a component of most parts of physics.
The Kramers-Kronig integrals are derived in a number of textbooks,\cite{jackson,LandauLifshitz,wooten,DresselGruner} and have been discussed by many authors.\cite{Toll56pr,Kubo72jmp,Hu89ajp,Peiponen91prb,Milton97prl,King02josab,King06jpa,Peiponen09rpp} The original derivations by Kramers and Kronig relied on model dielectric functions;\cite{Bohren10ejp}  however, the  subject is mostly approached by considering integrals on the complex frequency plane and using
Cauchy's integral theorem.\cite{Cauchy}  This approach, combined with the fact that the material's response functions are either even or odd as a function of the frequency and a consideration of the pole that occurs in conductors when the frequency $\omega$ is zero, lead to the following relations for the dielectric susceptibility $\chi = \chi_1 + i \chi_2$:
\begin{equation}
\chi_1(\omega) = {2 \over \pi}{{\cal{P}}}\int_{0}^{\infty} d\omega^{\prime}\, {\omega^{\prime} \chi_2(\omega^{\prime}) \over {\omega^{\prime}}^2 - \omega^2}, \label{KK4}
\end{equation}
and
\begin{equation}
\chi_2(\omega) =  {\sigma_{dc} \over \omega} -{2 \omega \over \pi} {\cal{P}}\int_{0}^{\infty} d\omega^{\prime}\,{\chi_1(\omega^{\prime}) \over {\omega^{\prime}}^2 - \omega^2}, 
\label{KK6}
\end{equation}
where $\sigma_{\rm dc}$ is the dc conductivity and  ${\cal{P}}$ means ``principal value.''

There are many complex optical functions in addition to the susceptibility: dielectric function, conductivity, refractive index, {\it etc\/}; there are Kramers-Kronig relations amongst   them. (Many can be obtained by substitution. For example, $\epsilon = 1 + 4\pi \chi$ and $\sigma_1 = \omega\chi_2 = \omega\epsilon_2/4\pi$ so that Equations \ref{KK4} and \ref{KK6} may respectively be converted to an integral containing $\sigma_1(\omega^{\prime})$ that gives $\epsilon_1(\omega)$ and one containing $\sigma_2(\omega')$ to give
$\sigma_1(\omega)$. Others require application of Cauchy's theorem to the contour integral along with statements about the behavior at very high frequencies.

\subsection{Kramers-Kronig analysis of reflectance}

When I measure the reflectance, where reflectance here means the {\it single-bounce\/} or {\it single-surface\/} reflectance, $\R$, I am taking the ratio  of the reflected intensity or power reflected from the front surface of the sample to the
incident intensity or power. The sample is assumed infinitely thick or sufficiently absorbing that no light from the rear surface reaches my detector.  Phase information is not available.

The amplitude reflectivity, the ratio of reflected electric field amplitude to incident electric field amplitude, does have a phase; indeed, I can write it as
\begin{equation}
r = \rho e^{i\phi} = {1 - N \over 1 + N}, \label{rvac}
\end{equation}
where $\rho = \sqrt{\R}$ is the magnitude of the reflectivity, $\phi$ is the phase shift on reflection, and $N = n+i \kappa$ is the complex refractive index. Here, the phase is set by measuring the field vector relative to the incident vector at the surface.\cite{branchcut}

It would be nice to know $\phi$, because I could invert Eq.~\ref{rvac} to get
\begin{equation}
N = { 1 - \sqrt{\R}e^{i\phi} \over 1 + \sqrt{\R}e^{i\phi} }, \label{Nfromr}
\end{equation}
using the known $\phi$  and the measured reflectance.

Kramers-Kronig analysis is one way of estimating this phase.\cite{wooten,Andermann65josa,Roessler65bjap}
Consider
\begin{equation}
\ln r = \ln \rho + i\phi.
\end{equation}
Here, $\ln \rho$ is the real part and  $ \phi$ is the imaginary part. The reflectance must be causal, and hence so must be the log of the reflectance. This requirement, plus the hermiticity of $r$, $r(-\omega) = [r(\omega)]^*$ (which makes $\rho$ even and $\phi$  odd) leads to
\begin{equation}
\phi(\omega) =  -{2 \omega \over \pi} {\cal{P}}\int_{0}^{\infty} d\omega^{\prime}\,{\ln \rho(\omega^{\prime}) \over {\omega^{\prime}}^2 - \omega^2}. \label{KK7}
\end{equation}

Equation \ref{KK7} is perfectly usable for numerical analysis, but there is an improvement that can be made.\cite{wooten} Consider
\begin{equation}
{\cal{P}}\int_{0}^{\infty}d\omega^{\prime} \,{1\over {\omega^{\prime}}^2 - \omega^2} = 0. 
\label{Int0}
\end{equation}
The   negative area for $\omega^{\prime}  < \omega$ cancels the positive area for
$\omega^{\prime} > \omega$. Thus, I can add
\begin{equation}
+ {2\omega \over \pi} \ln \rho(\omega) {\cal{P}}\int_{0}^{\infty} d\omega^{\prime}\,{1\over {\omega^{\prime}}^2 - \omega^2}
\end{equation}
to the right hand side of Eq.~\ref{KK7} without affecting the phase. Collecting terms, replacing $\rho$ with $\sqrt{\R}$, and using the properties of the log,
I get
\begin{equation}
\phi(\omega) =  -{ \omega \over \pi}  \int_{0}^{\infty}  d\omega^{\prime}\, {\ln [\R (\omega^{\prime}) /   \R (\omega)] \over {\omega^{\prime}}^2 - \omega^2}. \label{KKR}
\end{equation}

This modification has two advantages. First, if there are errors in the calibration of the reflectance measurements, so that the data for $\R$ are in error by a constant factor, the results for $\phi$ are unaffected. (Of course, even if the scale error does not affect the phase, it {\it does\/} affect $\R$, and $N$, Eq.~\ref{Nfromr}, depends on both quantities.) Second, both numerator and denominator of the integrand are zero when $\omega^{\prime} = \omega)$. L'H\^opital's rule shows that the ratio does not diverge; hence, the pole has been removed.

\subsection{Extrapolations}

The alert reader will have noticed that the range of the integral Eq.~\ref{KKR}  is 0 to $\infty$ and may wonder how one acquires data over that entire range. The answer is
that data are always limited to a finite range of frequencies. Thus, the user must use  extrapolations   outside the measured ranges.

\subsubsection{Low-frequency extrapolation}

One must estimate the reflectance between zero and  the lowest measured frequency. In my opinion, the best approach is to employ a model that reasonably describes the low-frequency data. Such models include Drude for metals, Lorentz for insulators, a sum of several Lorentzians, sometimes a Drude plus Lorentzians. Many other functions exist. When a good fit of the model to the data is obtained, a set of reflectance points may be calculated between zero and the lowest measured frequency, using a spacing between points similar to that of the lowest-frequency data, and combined with the measured data.

Other approaches to the low-frequency extrapolations include making an assumption that the reflectance is constant to DC (as might be appropriate for an insulator) or using a
Hagen-Rubens formula, $\R=1-A\omega^{1/2}$, (to describe a metal at low frequencies). Other power laws can also be used.
The constant $A$ is adjusted so the extrapolation goes through the first few points and then, using a spacing between points similar to that of the lowest-frequency data,
a set of reflectance points is  calculated between zero and the lowest measured frequency and combined with the measured data.

\subsubsection{High-frequency extrapolation}

The high-frequency extrapolation can be a source of major error. It is good to use data from other experiments on identical or similar samples if these exist. Moving to the highest frequencies, one knows that in the limit as $\omega\rightarrow \infty$, the dielectric function is mostly real and slightly smaller than unity, following $\epsilon = 1 - \omega_p^2/\omega^2$, where $\omega_p$ is the plasma frequency of all the electrons in the solid and $\omega\gg\omega_p$. Then $n  = 1 - \omega_p^2/2\omega^2$ and $\R = \omega_p^4/16\omega^4 \equiv C/\omega^4$.

Typically,\cite{wooten} the region between the highest-frequency data point and the transition to $C/\omega^{4}$ is filled with another power law, $\R \sim B/\omega^s$, with $  s  $ a free parameter. ($s$ does not have to be an integer.)  The value of $B$ is chosen so that the power law joins smoothly to the data at the high frequency limit and then     $C$ is   chosen for a smooth transition between mid- and high-frequency extrapolations. The free parameters are the exponent $s$ and the frequency for the crossover from $\omega^{-s}$ to $\omega^{-4}$. The integrals in the extrapolation range may be done analytically\cite{wooten}  and their contributions simply added to the  phase obtained by numerical integration of Eq.~\ref{KKR} over the low-frequency extrapolation and the measured data.

\section{Consequences of power law extrapolations}

\subsection{Silver metal}

\begin{figure}[tb]
\includegraphics[width=0.9\columnwidth] 
{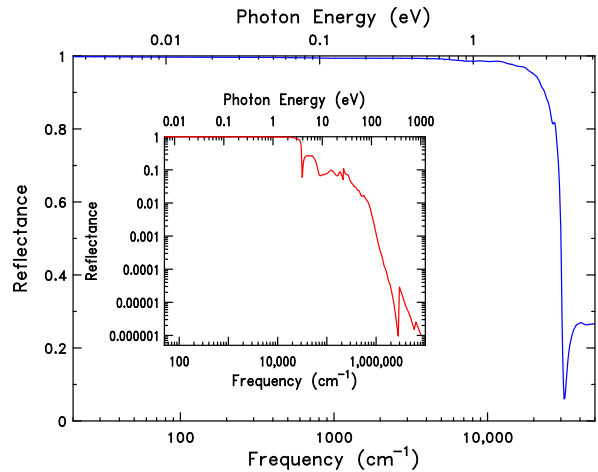}
\caption{(Color online)  Reflectance of silver.\cite{Palik,Taft61pr,Ehrenreich62pr,Bennett65ao,Palik14,Irani71josa,Hagemann75,Winsemius76,Leveque83}
The main panel shows the reflectance from far infrared to near ultraviolet whereas the inset shows the vacuum-ultraviolet and soft x-ray reflectance.}
\label{Fig:Ag-ins}
\end{figure}

An example of ``typical'' reflectance data   is shown in Fig.~\ref{Fig:Ag-ins}. The data are the reflectance of silver as collected by
Palik.\cite{Palik,Taft61pr,Ehrenreich62pr,Bennett65ao,Palik14,Irani71josa,Hagemann75,Winsemius76,Leveque83} The low frequencies are supplemented by a Drude reflectance, based on  resistivity $\rho=1.6~\mu\Omega$-cm at 300 K. The main panel shows the reflectance from 40--40,000 cm$^{-1}$ (5 meV--5 eV), a range that can be measured by many laboratories. One can see the high metallic reflectance from far-infrared to near ultraviolet, a sharp and deep plasma edge around 32,000 cm$^{-1}$ (4 eV), and the beginning of transitions from the d-bands to the conduction band above this.\cite{Ehrenreich62pr,wooten} The inset shows the data over the entire measured range, up to about 1 keV. 

I will now explore the Kramers-Kronig analysis of the reflectance shown in the main panel of Fig.~\ref{Fig:Ag-ins}. The issue to address is that the parameters, $s$ and the frequency of transition to $C/\omega^{4}$, are
completely free and, hence, uncontrolled. It is fair to ask: How much do they affect the outcome of the analysis? 
A reason for choosing silver for this discussion is that data for this material extend to 1000 eV (inset of Fig.~\ref{Fig:Ag-ins}). Here, one can see additional interband transitions followed by sharp core-level transitions. Note that the reflectance above 100 eV ($\sim 120,000$~cm$^{-1}$) is pretty close to a $\omega^{-4}$ power law. Kramers-Kronig analysis of the full data will be compared to the results of the limited data in the main panel.

\begin{figure}[tb]
\includegraphics[width=0.9\columnwidth] 
{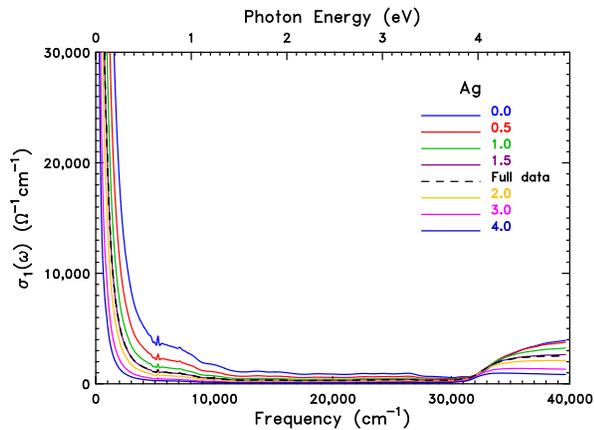}
\caption{(Color online)  Optical conductivity of silver from Kramers-Kronig analysis of reflectance. Power-law extensions were used, with exponents $0 < s < 4$ and a crossover to $\omega^{-4}$ at 10$^6$ cm$^{-1}$ (125 eV).}
\label{Fig:AgX}
\end{figure}

After Kramers-Kronig integration of the  reflectance, I can compute the optical conductivity, $\sigma_1(\omega)$, from the reflectance and phase. 
The results are shown in Fig.~\ref{Fig:AgX}. The intermediate frequency range was extrapolated as $\omega^{-s}$ with values for $s$ of 0, 0.5, 1.0, 1.5, 2, 3, and 4. The crossover to $\omega^{-4}$ occurred at 10$^6$ cm$^{-1}$ (125 eV).  There is considerable differences amongst the results. Note that all came to the same dc conductivity, $\sigma_{dc} = 620,000~\Omega^{-1}$cm$^{-1}$. The figure shows only the first or lowest 5\% of the  conductivity spectrum, to illustrate the variations in adequate detail.    The different power laws give a range of values for the conductivity in the d-transitions that  vary by a factor of three or so. Other optical function have similar variation. The conductivity from the full-range reflectance data is shown as the black dashed line, which covers the typical-range calculation for $s=1.5$. 

\begin{figure}[tb]
\includegraphics[width=0.9\columnwidth] 
{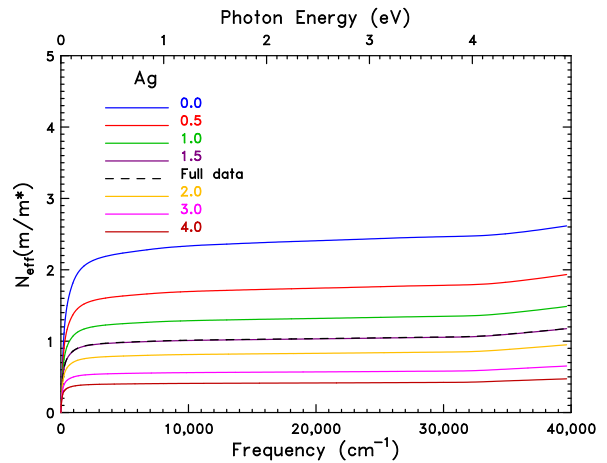}
\caption{(Color online) $(m /m^*)N_{{\it eff}}(\omega)$  of silver from Kramers-Kronig analysis of reflectance. Power-law extensions were used, with exponents $0 < s < 4$ and a crossover to $\omega^{-4}$ at 10$^6$0 cm$^{-1}$ (125 eV).}
\label{Fig:sumx}
\end{figure}

The extrapolation strongly affects the outcome for the partial sum rule for silver. This sum rule gives the number $N_{{\it eff}}$ of electrons with effective mass ratio $m/m^*$ participating in optical transitions at frequencies below $\omega$ as
\begin{equation}
{ m \over m^*}N_{{\it eff}}(\omega) =  {2m V_c \over   \pi e^2} \int_0^\omega \sigma_1(\omega') d\omega'
\label{partial}
\end{equation}
where $m^*$ is the effective mass,  $m$ is the free electron mass, $V_c$ is the unit cell volume (or formula volume).\cite{Vc} Figure \ref{Fig:sumx} shows the result of evaluating Eq.~\ref{partial} over 40--40,000 cm$^{-1}$ (5 mev--5 eV) for the conductivity data in Fig.~\ref{Fig:Ag-ins}. For a simple metal like silver, the free carrier spectral weight is exhausted in the midinfrared, and the function saturates at the number of conduction electrons/atom (1 in the case of Ag) until the interband transitions set in.\cite{Ehrenreich62pr,wooten} Depending on which power is taken in the intermediate region, this analysis would conclude that silver has between 0.4 and 2.4 free carriers per silver atom. Even if the experimenter avoided the extremes and chose $s$ in the range $1 \le s \le 3$, 
there would be a range of 0.6--1.3 for the number of electrons per silver atom. Note that the full-range data and $s = 1.5$ both saturate at $(m /m^*)N_{{\it eff}}(\omega)= 1.04$, a quite satisfactory result.\cite{Ehrenreich62pr}

\subsection{La$_{2-x}$Sr$_x$CuO$_4$}

\begin{figure}[tb]
\includegraphics[width=0.9\columnwidth] 
{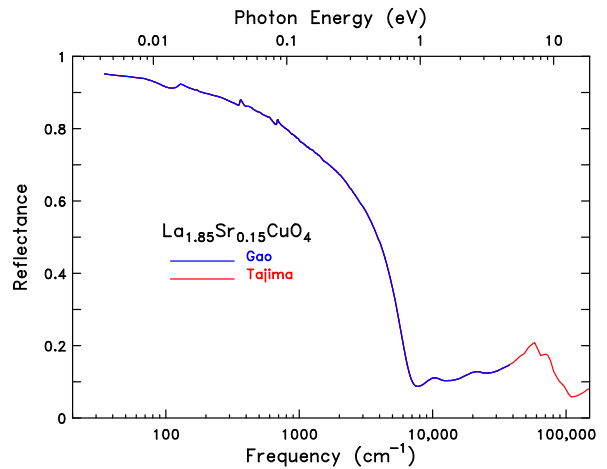}
\caption{(Color online) Reflectance (at 300 K) of La$_{1.85}$Sr$_{0.15}$CuO$_4$.}
\label{Fig:lsvuv}
\end{figure}

That this result is not unique to a free-carrier metal is evident when I repeat 
the exercise for La$_{1.85}$Sr$_{0.15}$CuO$_4$, using data from Gao et al.\cite{gao93prb} with ultraviolet results from Tajima et al.\cite{Tajima89josa} The data over the typical range\cite{gao93prb} cover 35--38,000 cm$^{-1}$; the ultraviolet results\cite{Tajima89josa} extend to 340,000 cm$^{-1}$ (42 eV). The reflectance data are shown in Fig.~\ref{Fig:lsvuv}. One can see a broad non-Drude midinfrared absorption with vibrational features superposed. The charge-transfer excitation of the insulator remains at 10,000 cm$^{-1}$ (1.2~eV) with higher-energy electronic transitions in the visible and ultraviolet.

\begin{figure}[tb]
\includegraphics[width=0.9\columnwidth]
{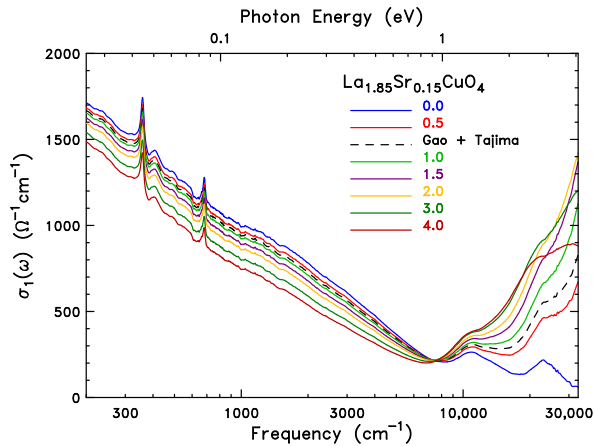}
\caption{(Color online) Optical conductivity (at 300 K) of La$_{1.85}$Sr$_{0.15}$CuO$_4$. Power-law extensions were used, with exponents $0 < s < 4$ and a crossover to $\omega^{-4}$ at 10$^6$ cm$^{-1}$ (125 eV). The conductivity obtained using the full data set, including the vacuum ultraviolet region,\cite{Tajima89josa} is also shown.} 
\label{Fig:lsold}
\end{figure}

The next step is to carry out the Kramers-Kronig integration of the limited-range data and use the phase so obtained to calculate the optical conductivity. Above the highest frequency of the measured data, the reflectance was extrapolated as $\omega^{-s}$ with values for $s$ of 0, 0.5, 1.0, 1.5, 2, 3, and 4. The crossover to $\omega^{-4}$ occurred at 10$^6$ cm$^{-1}$ (125 eV). The resulting optical conductivities are shown in Fig.~\ref{Fig:lsold}, along with a Kramers-Kronig-derived conductivity that includes the vacuum ultraviolet data.\cite{Tajima89josa}
Although all features appear, there is considerable variation in the spectral weights, particularly above about 1.2 eV (10,000 cm$^{-1}$).  The full-data spectrum falls midway between the results for $s=0.5$ and $s=1$. I find that the most worrisome feature is the large variation in the charge-transfer band, because one believes that the low-energy spectral weight is transferred from the charge-transfer spectrum of the insulator and would like to test this belief by measuring the spectral weight transfer.

\section{X-ray atomic scattering factors}

\subsection{Background}

Photoabsorption in the x-ray region has been considered by a number of
authors.\cite{Doyle68ac,Cromer68ac,Hubbell75,Henke81,Dreier84,henke88,Burge93,Chantler00}
It is described by an atomic scattering function, a complex quantity: $f = f_1 + i f_2$. 
The approach used is to combine experiment and theory and determine imaginary part of the scattering function, $f_2$, for each atomic species. This quantity has peaks or discontinuities at the absorption thresholds for each electronic level and falls to zero as $\omega \rightarrow  \infty$. The real part $f_1$ is obtained by Kramers-Kronig integration of $f_2$, It increases with frequency  via a series of plateaus, each approximately equal to the
number of ``free'' electrons at that photon energy, those electrons with  binding energies less than the photon energy. The limiting high frequency value is---except for relativistic corrections---then the atomic number $Z$.

In this work I use tables of the scattering functions reported by Henke, Gullikson, and  Davis\cite{henke93} about 20 years ago.  A related web site also exists\cite{henkeweb} with the ability to calculate optical properties, including reflectance. I find however that better results are found if two adjustments are made, one to the scattering functions $f$ and one to the procedure of using these function to calculate reflectance. The first is that the functions of Ref.~\onlinecite{henke93} provide $f_2$ from 10--30,000 eV but only have $f_1$ from 30--30,000 eV. I have redone the Kramers-Kronig integrals of $f_2$ to provide $f_1$ also from 10 eV (80,000 cm$^{-1}$), extrapolating  $f_2 \sim \omega^2$ at low frequencies and $f_2 \sim \omega^{-1}$
at high frequencies.

To obtain the optical properties of a material, one makes the assumption that the solid consists of a linear combination of its component atomic constituents, with the dielectric function determined by the scattering functions and the number density of the constituents. 
The dielectric function  is then
\begin{equation}
\epsilon = 1 - \sum_j{ 4 \pi n_j e^2\over m\omega^2} (f_{1}^j - i f_{2}^j),
\label{Henke}
\end{equation} 
where the sum runs over atoms $j$ at number density $n_j$ and with complex scattering function $f^j$. Note that this has the right limiting high-frequency behavior, because $f^j_1 \rightarrow  Z^j$ (with $Z^j$ here the atomic number) and $f_2 \rightarrow  0$, so that 
$\epsilon \rightarrow  1 - \sum_j{ 4 \pi n_j Z^j e^2 / m\omega^2}$

The complex refractive index is 
\begin{equation}
N = \sqrt{\epsilon} \label{Sqrt}
\end{equation} 
and the reflectance $\R$ is calculated from the usual equation, 
\begin{equation}
\R = \left|\frac{1 - N}{N + 1}\right|^2.  \label{Reflectance}
\end{equation}
Note that Ref.~\onlinecite{henke93} and the website, Ref.~\onlinecite{henkeweb}, write an equation for the refractive index of a monatomic solid:
\begin{equation} 
N = 1 - \frac{n r_0 \lambda^2}{2\pi} (f_1 - i f_2),
\end{equation} 
with $r_0 = e^2/mc^2$ the classical radius of the electron 
and $\lambda = 2\pi c/\omega$ the wavelength. This is clearly the first term in an expansion of $N = \sqrt{\epsilon}$. So  the second adjustment made for this work is to compute the dielectric function from $f$ using Eq.~\ref{Henke}, take the square root to obtain $N$, and then use Eq.~\ref{Reflectance} for the reflectance. 

Note also that many other sets of atomic scattering functions have been reported
\cite{henkeweb,Kane86pr,Sasaki89,Brennan92,Pratt94,Kissel95ac,Chantler95jpcrd,Cullen97}
in addition to the results in Ref.~\onlinecite{henke93}. In general the functions are similar at energies where they overlap; the newer sets often provide finer energy resolution near sharp features in the spectrum. One consequence is that, unlike the Henke functions, many are not sampled at the same photon energies, requiring the user to devise interpolation schemes when evaluating Eq.~\ref{Henke}.

\subsection{Implementation}

The procedure is implemented in the following way. The user supplies the chemical formula, such as Ag or La$_{1.85}$Sr$_{0.15}$CuO$_4$ and either  the appropriate\cite{Vc}  volume  $V_c$ or the density $\rho$. With this information, the reflectance $\R$ will be calculated at 340 logarithmically spaced points over 80,000--2.4$\times$10$^8$ cm$^{-1}$ (10-30,000 eV) using 
Eqs.~\ref{Henke}--\ref{Reflectance}.

A bridge need to be placed over the gap between the highest experimental point (say, 40,000 cm$^{-1}$ or 5 eV) and the beginning of the extrapolated reflectance at 80,000 cm$^{-1}$. The user has the option of a power series in $\omega$, in $1/\omega \propto \lambda$ or a cubic spline.\cite{spline}  As it turns out, the bridge has a modest effect in some cases, minimal in others.

A low-frequency extrapolation, from, say, 40 cm$^{-1}$ (5 meV) to zero, must be added. I find it effective to fit accurately the low-frequency reflectance to a Drude-Lorentz or other (well-motivated) model and calculate the low-frequency reflectance from the model.

Then, the Kramers-Kronig integral, Eq.~\ref{KKR}, is computed to obtain the phase. The refractive index can then be calculated from reflectance and phase through Eq.~\ref{Nfromr}, with other optical constants following in the usual way. 

\subsection{Silver metal}

\begin{figure}[tb]
\includegraphics[width=0.9\columnwidth]
{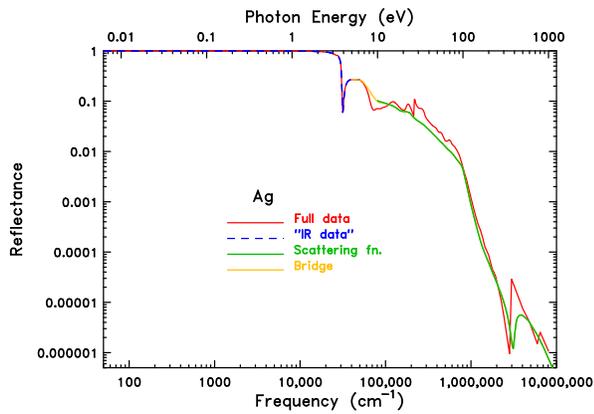}
\caption{(Color online)  Reflectance of silver over 1--$10^7$ cm$^{-1}$ (1--1200 eV) showing the typical experimental region (40--40,000 cm$^{-1}$), data in the vacuum-ultraviolet and soft x-ray region\cite{Palik}, the reflectance calculated from the Ag scattering function, and the power-law bridge over 50,000--80,000 cm$^{-1}$ (5--20eV).}
\label{Fig:Agbr}
\end{figure}

To start, I will explore the use of the atomic scattering functions to analyze the reflectance data   of Fig.~\ref{Fig:Ag-ins} (main panel). The scattering function $f$ for silver is used to calculate the reflectance in the UV--X-ray region. The beginning of the calculated X-ray reflectance is shown in Fig.~\ref{Fig:Agbr}. (It extends to 30,000 eV, continuing to fall approximately as $1/\omega^4$ with some fine structure.) 
The measured high frequency reflectance\cite{Palik} also is shown, along with a power-law bridge between the infrared--UV data and the beginning of the X-ray calculation. The calculated reflectance follows experiment reasonably well, although it is higher at the beginning and lower in the 2$\times$10$^5$--$10^6$~cm$^{-1}$ (30--100 eV) region. The strong structure around 3$\times$10$^6$ cm$^{-1}$ (400 eV) appears, although broader than experiment.

\begin{figure}[tb]
\includegraphics[width=0.9\columnwidth] 
{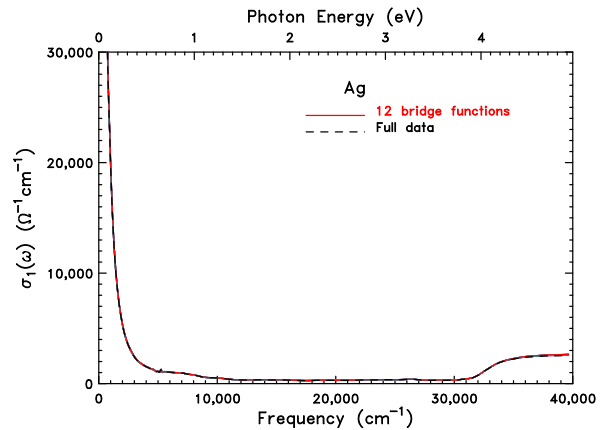}
\caption{(Color online) Optical conductivity of silver from Kramers-Kronig analysis of reflectance. An X-ray scattering-function extrapolation with 12 different bridge function (described in the text) was used. The conductivity from Kramers-Kronig analysis of the full data set is also shown.}
\label{Fig:s1br}
\end{figure}

My goal, of course, is to use the scattering function reflectance for Kramers-Kronig extension, not to extract accurate UV--X-ray reflectance. The conductivity obtained by Kramers-Kronig analysis using the scattering function extension is shown in Fig.~\ref{Fig:s1br}. Twelve curves are shown for twelve variations of the bridge function: power laws of $\sum_0^n a_n \omega^n$ with $n$ varying from 1 (straight line) to 5 and power laws of   $\sum_0^n b_n \omega^{-n}$ with $n$ varying from 1   to 7. Finally, a curve for the full data is also shown. There is no significant difference among the results. All of the bridge functions give conductivity spectra basically indistinguishable from the result from the full data. There is a small deviation in a couple of cases above 35,000 cm$^{-1}$, which I regard as not really significant.

\begin{figure}[tb]
\includegraphics[width=0.9\columnwidth] 
{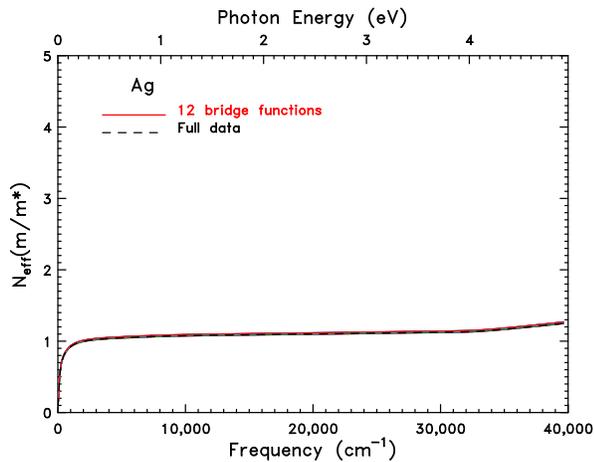}
\caption{(Color online)  $(m /m^*)N_{{\it eff}}(\omega)$  of silver from Kramers-Kronig analysis of reflectance using  X-ray scattering-function extrapolation with 12 different bridge function (described in the text). The partial sum rule for the full data is also shown.}
\label{Fig:sumbr}
\end{figure}

Fig.~\ref{Fig:sumbr} shows that the result for the partial sum rule is equally good. Any of the bridge functions would support the notion that there is about 1 free electron per silver atom.

\subsection{La$_{2-x}$Sr$_x$CuO$_4$}

\begin{figure}[tb]
\includegraphics[width=0.9\columnwidth]
{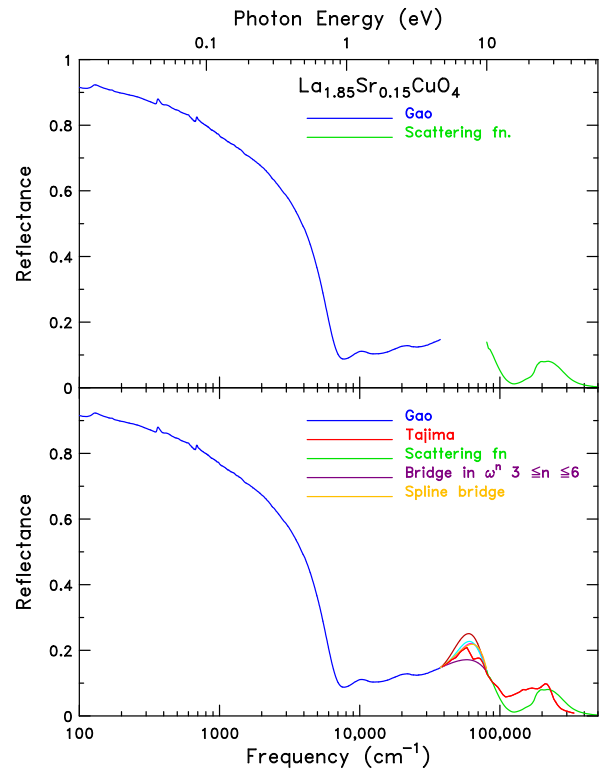}
\caption{(Color online)  Reflectance (at 300 K) of La$_{1.85}$Sr$_{0.15}$CuO$_4$. The upper panel shows the data of Gao et al.\cite{gao93prb} and the reflectance calculated from the X-ray scattering functions. The bottom panel shows the data and scattering-function reflectance again, along with 5 bridges, as described in the text. The actual UV reflectance\cite{Tajima89josa} is also shown.}
\label{Fig:lsxro}
\end{figure}

That this result is not unique to a free-carrier metal is evident when I repeat 
the exercise for La$_{1.85}$Sr$_{0.15}$CuO$_4$.\cite{gao93prb,Tajima89josa} This material is a good test of the method, because, as shown in the upper panel of Fig.~\ref{Fig:lsxro}, the slopes of the data at the high-frequency limit, which is about 37,000 cm$^{-1}$ (4.6 eV), is positive whereas the slope of the scattering-function reflectance at its low-frequency limit, which is about 80,000 cm$^{-1}$ (10 eV), is negative. Hence the bridge must provide this slope change.

\begin{figure}[tb]
\includegraphics[width=0.9\columnwidth] 
{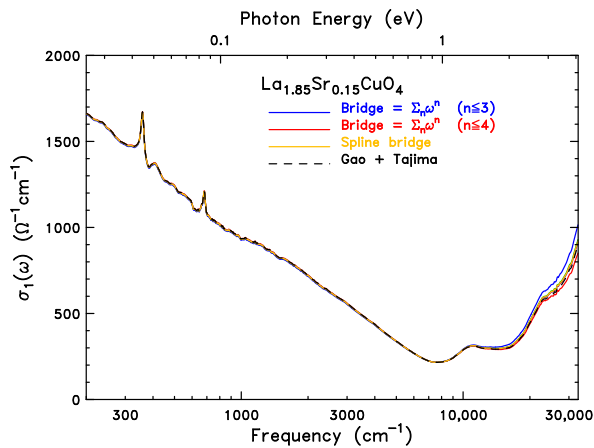}
\caption{(Color online) Optical conductivity (at 300 K) of La$_{1.85}$Sr$_{0.15}$CuO$_4$. Scattering-function extensions were used, with a variety of power-law in $\omega$ bridge functions spanning the gap between measured and scattering-function extension.  The conductivity obtained using the full data set, including the vacuum ultraviolet region,\cite{Tajima89josa} is also shown.} 
\label{Fig:lsnew2}
\end{figure}

Several trial bridge functions are shown in the bottom panel. 
Four used power laws in $\omega$, $\R = \sum_0^n A_n\omega^n$ with the upper limit ranging from 3 to 6. A bridge employing a cubic spline function is also shown. All accomplish the goal of joining the two regions, with some above and some below the the actual UV reflectance,\cite{Tajima89josa} which also is shown. 
The agreement between the scattering-function-derived reflectance and the actual reflectance is not as good as in the case of Ag. However, I am not that interested in the accuracy of the scattering-function reflectance in the vacuum UV and X-ray region; instead I will use it as an extension in the Kramers-Kronig analysis of infrared--UV reflectance. The outcome of the Kramers-Kronig analysis is shown in Fig.~\ref{Fig:lsnew2}. Above the highest measured or calculated point, the reflectance was extended as $1/\omega^4$. Below the lowest measured frequency, it was extended with a Drude-Lorentz fit.

As opposed to the results of using power-law extensions in the intermediate region, which were shown in Fig.~\ref{Fig:lsold}, the scattering-function extension is quite close to the conductivity obtained using actual vacuum-ultraviolet data. In particular, the conductivity through the important charge-transfer band around 12,000 cm$^{-1}$ (1.5 eV) is almost independent of bridge function and is very close to what is found using actual data. Evaluation of the partial sum rule, Eq.~\ref{partial}, works the same way; the curves are nearly indistinguishable below about 16,000 cm$^{-1}$ (2 eV). Of the various bridge functions, the cubic spline\cite{spline} appears to be closest to the result using the full data (Gao + Tajima).\cite{gao93prb,Tajima89josa} The two extremes used cubic (high) and quartic (low) power laws in $\omega$; carrying the series to more terms made little difference  and neither did the use of power laws in $1/\omega$.

\subsection{Aluminum}

\begin{figure}[tb]
\includegraphics[width=0.9\columnwidth] 
{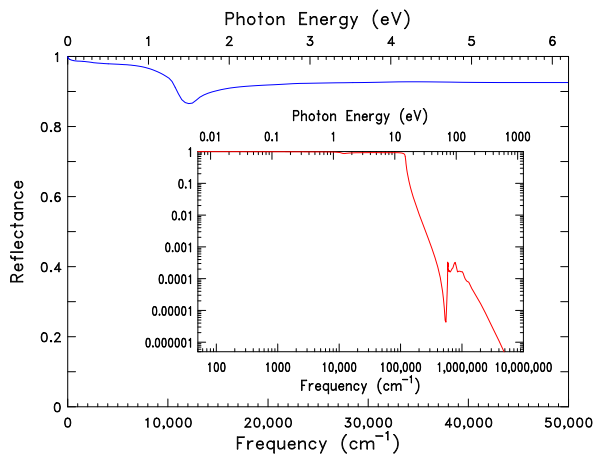}
\caption{(Color online)  Reflectance of aluminum.\cite{Palik,Ehrenreich63pr,Hagemann75,Shiles80prb,Smith86prb}
The main panel shows the reflectance from far infrared to near ultraviolet whereas the inset shows the vacuum-ultraviolet and soft x-ray reflectance.}
\label{Fig:Al-ins}
\end{figure}

\begin{figure}[tb]
\includegraphics[width=0.9\columnwidth] 
{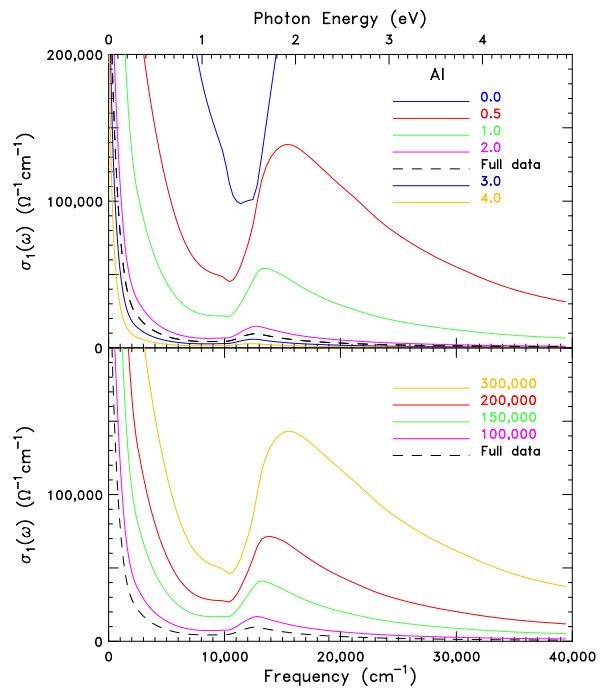}
\caption{(Color online) Optical conductivity of aluminum from Kramers-Kronig analysis of reflectance. In the upper panel, power-law extensions were used, with exponents $0 < s < 4$ and a crossover to $\R \sim \omega^{-4}$ at 10$^6$ cm$^{-1}$ (125 eV). In the lower panel, a constant reflectance was used, up to a changeover frequency $100,000 < \omega_a < 500,000$ cm$^{-1}$, above which $\R \sim \omega^{-4}$.}
\label{Fig:alold}
\end{figure}

As a final example, let me discuss the use of this method for Al metal. A trivalent metal, Al has its plasma edge deep in the vacuum ultraviolet, around 120,000 cm$^{-1}$ (15 eV). This energy is well beyond the reach of most conventional optical spectroscopy laboratories. Nevertheless, one might study Al (or similar wide-band, 
high-carrier-density solids) in order to probe low-energy features, such as the weak interband transition that occurs in the infrared.\cite{Brust70prb,Smith86prb} The far-infrared--ultraviolet reflectance spectrum\cite{Palik,Ehrenreich63pr,Hagemann75,Shiles80prb,Smith86prb} of Al is shown in Fig.~\ref{Fig:Al-ins}. The main panel shows data from 70--50,000 cm$^{-1}$ (0.01--6.2 eV) whereas the inset shows the vacuum ultraviolet and X-ray reflectance up to 6$\times$10$^6$~cm$^{-1}$ (800 eV). Were I to have only the low energy data of the main panel, I would have a difficult time with the Kramers-Kronig analysis. I would suspect that the high reflectance did not continue indefinitely but would not {\it a priori\/} know when or how to start it decreasing. If I were to use 
a power law for the reflectance, $\R \sim 1/\omega^n$, transitioning to $\R \sim 1/\omega^{4}$ at a considerably higher frequency and continuing this free-carrier extension to infinity, I would obtain the curves  shown in Fig.~\ref{Fig:alold}.

The upper panel shows the effect of varying the exponent in the mid-region from 0 (a nearly flat extension, as suggested by the data) to 4. The transition to $\R \sim 1/\omega^4$ was made at 10$^6$ cm$^{-1}$ (125 eV). The Kramers-Kronig-derived conductivity from the full data set (shown in the inset to Fig.~\ref{Fig:Al-ins}) is also shown. Most of the derived curves seriously overestimate conductivity, making the spectral weight in both the free-carrier conductivity and the weak interband transition far too large. The location of the maximum of this transition is pushed up in the small exponent calculations, reaching 16,000 cm$^{-1}$ (2 eV) compared to 12,500 cm$^{-1}$ (1.55 eV) in the conductivity derived from the full-range-data. Note that the weak exponents (which are implied by the data) give the poorest results. An exponent of around 2.5, which is in no way suggested by the data, gives a conductivity spectrum close to that returned by using the full reflectance spectrum.

One could argue that transitioning to $\R \sim \omega^{-4}$ at 10$^6$ cm$^{-1}$ (125 eV) is not correct; the full reflectance spectrum turns over at 120,000 cm$^{-1}$. The lower panel of Fig.~\ref{Fig:alold} shows  conductivity spectra obtained by extending the reflectance as a constant value up to a frequency $\omega_a$ where it changes to $\R \sim \omega^{-4}$. The changeover frequency was in the range
 $100,000 < \omega_a < 500,000$ cm$^{-1}$ (12--60~eV); above this frequency $\R \sim \omega^{-4}$. 
 All of the curves overstate the magnitude of the conductivity, even the one where the reflectance decrease starts at 100,000 cm$^{-1}$ (12 eV),  below the 150,000 cm$^{-1}$ (19 eV)  where the experimental edge exists. Note that the initial power-law behavior of the reflectance edge (inset to Fig.~\ref{Fig:Al-ins}) is approximately $\omega^{-9}$. 
 
\begin{figure}[tb]
\includegraphics[width=0.9\columnwidth]  
{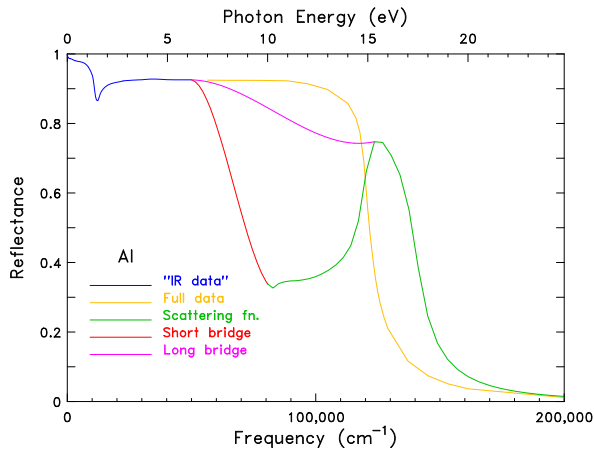}
\caption{(Color online) Reflectance of aluminum over 1--200,000 cm$^{-1}$ (1--25 eV), showing the typical experimental region (40--50,000 cm$^{-1}$), data in the vacuum-ultraviolet and soft x-ray regoion\cite{Palik}, the calculated reflectance using the Al scattering function, and two cubic spline  bridges, one over 50,000--80,000 cm$^{-1}$ (6.2--10 eV) and one over 50,000--130,000 cm$^{-1}$ (6.2--16 eV).}
\label{Fig:alxro3}
\end{figure}

 Presented with the spectra in Fig.~\ref{Fig:alold}, I would decide that Kramers-Kronig analysis of the reflectance in the main panel of Fig.~\ref{Fig:Al-ins} cannot be productive. However, use of the X-ray scattering function helps immensely. The relevant curves are shown in Fig.~\ref{Fig:alxro3}.

\begin{figure}[tb]
\includegraphics[width=0.9\columnwidth] 
{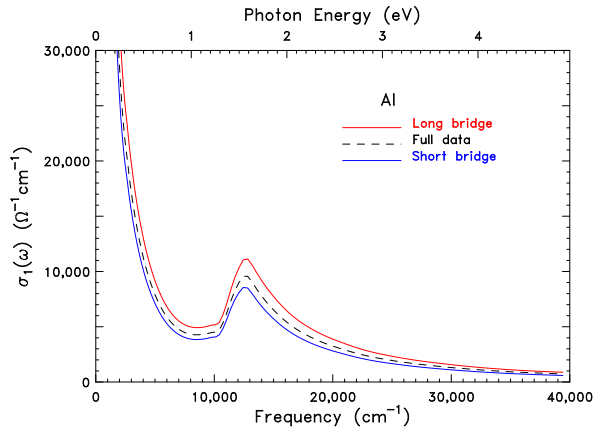}
\caption{(Color online)  Optical conductivity of aluminum from Kramers-Kronig analysis of reflectance. An X-ray scattering function extrapolation with cubic spline bridge functions (described in the text) was used. The conductivity from Kramers-Kronig analysis of the full data set is also shown.}
\label{Fig:AlS4br}
\end{figure}

At first blush, the calculated reflectance from the scattering functions does not look promising. Because there is no band structure, only atomic orbitals, there is no metallic reflectance; instead a strong peak  occurs around 130,000 cm$^{-1}$ (16 eV) with a reflectance edge blue shifted from the experimental data. At the low end, 80,000 cm$^{-1}$ (10 eV), the reflectance falls to about 1/3 of the metal's reflectance. I consider two approaches to the use of this extension. First, I can just use it, with a short, steeply declining bridge joining the ``IR data'' to the 
scattering-function extension. Second, I can lop off the low energy part of the extension, and bridge to the maximum around 130,000 cm$^{-1}$ (16 eV), making the smallest change in reflectance between data and extension. Surprisingly, both approaches give rather similar optical conductivity curves (and sum rule results), as shown in Fig.~\ref{Fig:AlS4br}.

The short bridge falls a little bit below the conductivity obtained from the full set of data while the long bridge is a bit above. I could use either (or their average!) to discuss the low frequency electronic structure of aluminum without being plagued by extrapolation-dependent results. That the scattering-function extrapolation works as well as it does suggests that the critical issue in designing the extrapolation is to use one that  gets the correct high-energy spectral weight for the material and then places that spectral weight appropriately in energy. The remaining details   are not important.

\section{Conclusions}

The use of reflectance calculated from a dielectric function constructed from a sum of atomic scattering functions for a material provides a reliable and reproducible method of extrapolating measured data. It removes a certain amount of arbitrariness in the use of Kramers-Kronig analysis.

In addition to testing with data where the UV--X-ray spectra are known, this extrapolation has been  used in a number of recent 
studies.\cite{Miller13prb,Turner14,Anand14,Martin14} A comparison with a method that uses ellipsometry in the near-infrared--ultraviolet to constrain the extrapolation\cite{Stricker14prl}  gave conductivity spectra indistinguishable from those of  Figs.~\ref{Fig:s1br} and \ref{Fig:lsnew2}. For Al, the ellipsometry-based method gave a slightly (5\%) higher conductivity over the entire range; the difference was about the same as that between the two bridge versions in Fig.~\ref{Fig:alxro3}.

Persons who wish to test the approach will find a Windows program to compute the reflectance in the UV--X-ray region, the Kramers-Kronig routine that uses the extrapolation and generates the bridge function, and a program that computes optical constants from reflectance and phase at \path |http://www.phys.ufl.edu/~tanner/ZIPS/xro.zip|. Use {\tt xro.exe} to generate the file for extrapolation, {\tt kk.exe} to do the Kramers-Kronig integral, and {\tt op.exe} to calculate optical functions. Note that to avoid having $2\pi$ swings in the phase when its value is close to 
$\phi = -\pi$ and there is noise, the {\tt kk.exe} and {\tt op.exe} programs respectively  compute and use $\theta = \phi + \pi$. See the discussion in endnote~\onlinecite{branchcut}.

\begin{acknowledgments}
 I've had important discussion about this method with Ric Lobo,  Tom Timusk, and Dirk ven der Marel. Claus Jacobsen wrote the Kramers-Kronig routine I use and Charles Porter wrote many of the  routines used by the data analysis code. Naween Anand, Chang Long, Catalin Martin, Kevin Miller, Zahra Nasrollahi, Evan Thatcher,  
Berik Uzakbaiuly, and  Luyi Yan were very helpful in testing  the method on data they have measured. I thank Dirk van der Marel and Damien Stricker for providing a comparison of the method used here with their ellipsometry-based approach.
\end{acknowledgments}


\begin{thebibliography}{99}


\bibitem {Kronig} R. de L. Kronig, ``On the theory of the dispersion of X-rays,'' \journal J. Opt. Soc. Am., 12, 547--557, 1926 \par 

\bibitem {Kramers} H.A. Kramers, ``La diffusion de la lumiere par les atomes,''  J. Atti Cong. Intern. Fisici, (Transactions of Volta Centenary Congress) Como {\bf 2}, 545--557 (1927); ``Die dispersion und absorption von R\"ontgenstrahlen,''  Phys. Z. {\bf 30}, 522--523 (1929).

\bibitem {jackson} J.D. Jackson, {\it Classical Electrodynamics\/} (John Wiley \& Sons, New York, 1975).   \par 

\bibitem {LandauLifshitz} L.D. Landau and E.M. Lifshitz, {\it Electrodynamics of Continuous Media\/} (Pergamon, New York, 1960). \par 

\bibitem {wooten} Frederick Wooten, {\it Optical Properties of Solids\/} (Academic Press, New York, 1972). \par 

\bibitem {DresselGruner} Martin Dressel and George Gr\"uner, {\it Electrodynamics of Solids: Optical Properties of Electrons in Matter\/} (Cambridge University Press, Cambridge, 2002). \par 

\bibitem {Jahod57pr} Franz C. Jahod, ``Fundamental Absorption of Barium Oxide from Its Reflectivity Spectrum,'' Phys. Rev. {\bf 107}, 1261--1265 (1957). \par 

\bibitem {Taft61pr} E.A. Taft and H.R. Philipp, ``Optical Constants of Silver,'' Phys. Rev. {\bf 121}, 1100--1103 (1961). \par 

\bibitem {Ehrenreich62pr} H. Ehrenreich and H.R. Philipp, ``Optical Properties of Ag and Cu'' Phys. Rev. {\bf 128}, 1622 (1962). \par 

\bibitem {Ehrenreich63pr} H. Ehrenreich, H.R. Philipp, and B. Segall, ``Optical Properties of Aluminum,'' Phys. Rev. {\bf 132}, 1918--1928 (1963). \par 

\bibitem {Irani71josa} G. Irani, T. Huen, and F. Wooten, ``Optical Constants of Silver and Gold in the Visible and Vacuum Ultraviolet,'' J. Opt. Soc. Am. {\bf 61}, 128--129 (1971). \par 

\bibitem {Hagemann75} H.J. Hagemann, W. Gudat, and C. Kunz, DESY Report SR-74/7, Hamburg (1974); H.J. Hagemann, W. Gudat, and C. Kunz, ``Optical constants from the far infrared to the x-ray region: Mg, Al, Cu, Ag, Au, Bi, C, and Al$_2$O$_3$,'' J. Opt. Soc. Am. {\bf 65}, 742 (1975). \par 

\bibitem {Winsemius76} P. Winsemius, F.F. van Kampen, H.P. Lengkeek, and C.G. van Went, ``Temperature dependence of the optical properties of Au, Ag, and Cu,'' J. Phys. F {\bf 6}, 1583 (1976). \par 

\bibitem {Shiles80prb} E. Shiles, Taizo Sasaki, Mitio Inokuti, and D.Y. Smith, ``Self-consistency and sum-rule tests in the Kramers-Kronig analysis of optical data: Applications to aluminum,'' Phys. Rev. B {\bf 22}, 1612 (1980). \par 

\bibitem {Leveque83} G. Leveque, C.G. Olson, and D.W. Lynch, ``Reflectance spectra and dielectric functions for Ag in the region of interband transitions,'' Phys. Rev. B {\bf 27}, 4654 (1983). \par 

\bibitem {Phillip64pr} H.R. Phillip and E.A. Taft, ``Kramers-Kronig Analysis of Reflectance Data for Diamond,'' Phys. Rev. {\bf 136}, A1445--A1448 (1964). \par 

\bibitem {Taft65pr} E.A. Taft and H.R. Philipp, ``Optical Properties of Graphite,'' Phys. Rev. {\bf 138}, A197--A202 (1965). \par 

\bibitem {Tongay10prb} S. Tongay, J. Hwang, D.B. Tanner, H.K. Pal, D. Maslov, and A.F. Hebard, ``Supermetallic conductivity in bromine-intercalated graphite,'' \PRB 81, 115428, 2010 \par 

\bibitem {Jacobsen74prl} C.S. Jacobsen, D.B. Tanner, A.F. Garito, and A.J. Heeger, ``Single-crystal reflectance studies of tetrathiofulvalene tetracyanoquinodimethane,'' \PRL 33, 1559--1562, 1974 \par 

\bibitem {Jacobsen81prl} C.S. Jacobsen, D.B. Tanner, and K. Bechgaard, ``Dimensionality crossover in the organic superconductor tetramethyltetraselenafulvalene hexafluorophosphate [(TMTSF)$_2$PF$_6$],'' \PRL 46, 1142--1145, 1981 \par 

\bibitem {Dressel96prl} M. Dressel, A. Schwartz, G. Gr\"uner, and L. Degiorgi, ``Deviations from Drude Response in Low-Dimensional Metals: Electrodynamics of the Metallic State of (TMTSF)$_2$PF$_6$,'' Phys. Rev. Lett. {\bf 77}, 398--402 (1996). \par 

\bibitem {Beal79jpc} A.R. Beal and H.P. Hughes, ``Kramers-Kronig analysis of the reflectivity spectra of 2H-MoS$_2$, 2H-MoSe$_2$ and 2H-MoTe$_2$,'' J. Phys. C: Solid State Phys. {\bf 12}, 881 (1979) \par 

\bibitem {Greene82prb} L.H. Greene, D.B. Tanner, A.J. Epstein, and Joel S. Miller, ``Optical properties of the cation-deficient platinum chain salt
  K$_{1.75}$Pt(CN)$_4\cdot 1.5$H$_2$O,'' \PRB 25, 1331--1338, 1982 \par 

\bibitem {Challener84ssc} W.A. Challener and P.L. Richards, ``Far infrared optical properties of NbSe$_3$,'' Solid State Comm. { \bf 52}, 117--121 (1984). \par 

\bibitem {Fincher79prb} C.R. Fincher, Jr., M. Ozaki, M. Tanaka, D. Peebles, L. Lauchlan, A.J. Heeger, and A.G. MacDiarmid, ``Electronic structure of polyacetylene: Optical and infrared studies of undoped semiconducting (CH)$_x$ and heavily doped metallic (CH)$_x$,'' Phys. Rev. B {\bf 20}, 1589 (1979). \par 

\bibitem {Stafstrom87prl} S. Stafstr\"om, J.L. Br\'edas, A.J. Epstein, H.S. Woo, D.B. Tanner, W.S. Huang, and A.G. MacDiarmid, ``Polaron lattice in highly conducting polyaniline: Theoretical and optical studies,'' \PRL 59, 1464--1467, 1987 \par 

\bibitem {Lee06nat} K. Lee, S. Cho, S.H. Park, A.J. Heeger, C.-W. Lee, and S.-H. Lee, ``Metallic transport in polyaniline,'' Nature {\bf 441}, 65--68 (2006). \par 

\bibitem {Bonn87prl} D.A. Bonn, J.E. Greedan, C.V. Stager, T. Timusk, M.G. Doss, S.L. Herr, K. Kamar\'as, and D.B. Tanner, ``Far-infrared conductivity of the high-$T_c$ superconductor  YBa$_2$Cu$_3$O$_7$,'' \PRL 58, 2249--2250, 1987 \par 

\bibitem {Reedyk88prb} M. Reedyk, D.A. Bonn, J.D. Garrett, J.E. Greedan, C.V. Stager, T. Timusk, K. Kamar\'as, and D.B. Tanner, ``Far-infrared optical properties of $\rm Bi_2Sr_2CaCu_2O_8$,'' \PRB 38, {11\,981--11\,984}, 1988 \par 

\bibitem {Kamaras90prl} K. Kamar\'as, S.L. Herr, C.D. Porter, N. Tache, D.B. Tanner, S. Etemad, T. Venkatesan, E. Chase, A. Inam, X.D. Wu, M.S. Hegde, and B. Dutta, ``In a clean high-$T_c$ superconductor you do not see the gap,'' \PRL 64, 84--87, 1990 \par 

\bibitem {cooper90prb1} S.L. Cooper, G.A. Thomas, J. Orenstein, D.H. Rapkine, A.J. Millis, S.-W. Cheong, A.S. Cooper, and Z. Fisk, ``Growth of the optical conductivity in the Cu-O planes,'' \PRB 41, {11605--11608}, 1990 \par 

\bibitem {gao93prb} F. Gao, D.B. Romero, D.B. Tanner, J. Talvacchio, and M.G. Forrester, ``Infrared properties of epitaxial La$_{2-x}$Sr$_{x}$CuO$_4$ thin films in the normal and superconducting states,'' \PRB 47, 1036--1052, 1993 \par 

\bibitem {cooper93prb} S.L. Cooper, D. Reznik, A. Kotz, M.A. Karlow, R. Liu, M.V. Klein, W.C. Lee, J. Giapintzakis, D.M. Ginsberg, B.W. Veal, and A.P. Paulikas, ``Optical studies of the $a$-, $b$-, and $c$-axis charge dynamics in YBa$_{2}$Cu$_{3}$O$_{6+x}$,'' \PRB 47, 8233--8248, 1993 \par 

\bibitem {homes93prl} C.C. Homes, T. Timusk, R. Liang, D.A. Bonn, and W.N. Hardy, ``Optical conductivity of c axis oriented YBa$_2$Cu$_3$O$_{6.70}$: Evidence for a pseudogap,'' \PRL 71, 1645, 1993   \par 

\bibitem {Basov94prb2} D.N. Basov, A.V. Puchkov, R.A. Hughes, T. Strach, J. Preston, T. Timusk, D.A. Bonn, R. Liang, and W.N. Hardy, ``Disorder and superconducting-state conductivity of single crystals of YBa$_2$Cu$_3$O$_{6.95}$,'' Phys. Rev. B {\bf 49}, 12165--12169 (1994).  \par 

\bibitem {Basov95prl} D.N.~Basov, R.~Liang, D.A.~Bonn, W.N. Hardy, B.~Dabrowski, M.~Quijada, D.B.~Tanner, J.P.~Rice, D.M.~Ginsberg, and T.~Timusk, ``In-plane anisotropy of the penetration depth in YBa$_2$Cu$_3$O$_{7-x}$ and YBa$_2$Cu$_4$O$_8$ superconductors,'' \PRL 74, 598--601, 1995 \par 

\bibitem {Quijada99prb} M.A. Quijada, D.B. Tanner, R.J. Kelley, M. Onellion, H. Berger, and G. Margaritondo ``Anisotropy in the $ab$-plane optical properties of  Bi$_2$Sr$_2$CaCu$_2$O$_7$
single-domain crystals,'' \PRB 60, 14\,917--14\,934, 1999 \par 

\bibitem {Puchkov96prl} A.V. Puchkov, P. Fournier, T. Timusk, and N.N. Kolesnikov, ``Optical Conductivity of High Tc Superconductors: From Underdoped to Overdoped,'' Phys. Rev. Lett. {\bf 77}, 1853 (1996). \par 

\bibitem {Kaplan96prl} S.G. Kaplan, M.A. Quijada, H.D. Drew, D.B. Tanner, G.C. Xiong, R. Ramesh, C. Kwon, and T. Venkatesan, ``Optical evidence for the dynamic Jahn-Teller effect in Nd$_{0.7}$Sr$_{0.3}$MnO$_3$,'' \PRL 77, 2081-2084, 1996 \par 

\bibitem {Murakami98prl} Y. Murakami, H. Kawada, H. Kawata, M. Tanaka, T. Arima, Y. Moritomo, and Y. Tokura,
 ``Direct Observation of Charge and Orbital Ordering in La$_{0.5}$Sr$_{1.5}$MnO$_{4}$,'' Phys. Rev. Lett. {\bf 80}, 1932--1935 (1998). \par 

\bibitem {Kim98prl} K.H. Kim, J.H. Jung, and T.W. Noh, ``Polaron Absorption in a Perovskite Manganite La$_{0.7}$Ca$_{0.3}$MnO$_{3}$,'' Phys. Rev. Lett. {\bf 81}, 1517--1520 (1998). \par 

\bibitem {Li08prl} G. Li, W.Z. Hu, J. Dong, Z. Li, P. Zheng, G.F. Chen, J.L. Luo, and N.L. Wang, ``Probing the Superconducting Energy Gap from Infrared Spectroscopy on a Ba$_{0.6}$K$_{0.4}$Fe$_{2}$ As$_{2}$ Single Crystal with $T_c=37$~K,'' Phys. Rev. Lett. {\bf 101}, 107004 (2008). \par 

\bibitem {Lucarelli10njp} A. Lucarelli, A. Dusza, F. Pfuner, P. Lerch, J.G. Analytis, J.-H. Chu, I.R. Fisher, and L. Degiorgi, ``Charge dynamics of Co-doped BaFe$_2$As$_2$,'' New J. Phys. {\bf 12}, 073036 (2010). \par 

\bibitem {Barisic10prb} N. Bari\v {s}i\'c, D. Wu, M. Dressel, L. J. Li, G. H. Cao, and Z. A. Xu, ``Electrodynamics of electron-doped iron pnictide superconductors: Normal-state properties,'' Phys. Rev. B {\bf 82}, 054518 (2010). \par 

\bibitem {Cheng12prb} B. Cheng, B.F. Hu, R.Y. Chen, G. Xu, P. Zheng, J.L. Luo, and N.L. Wang, ``Electronic properties of 3d transitional metal pnictides: A comparative study by optical spectroscopy,'' Phys. Rev. B {\bf 86}, 134503 (2012). \par 

\bibitem {Dai12prb} Y.M. Dai, B. Xu, B. Shen, H.H. Wen, J.P. Hu, X.G. Qiu, and R.P.S.M. Lobo, ``Pseudogap in underdoped Ba$_{1-x}$K$_x$Fe$_2$As$_2$ as seen via optical conductivity,'' Phys. Rev. B {\bf 86}, 100501(R) (2012). \par 

\bibitem {Moon13prl} S.J. Moon, A.A. Schafgans, M.A. Tanatar, R. Prozorov, A. Thaler, P.C. Canfield, A.S. Sefat, D. Mandrus, and D.N. Basov, ``Interlayer Coherence and Superconducting Condensate in the $c$-Axis Response of Optimally Doped Ba(Fe$_{1-x}$Co$_x$)$_2$As$_2$ High-$T_c$ Superconductor Using Infrared Spectroscopy,'' Phys. Rev. Lett. {\bf 110}, 097003 (2013). \par 

\bibitem {Dai13epl} Y.M. Dai, B. Xu, B. She2, H.H. Wen, X.G. Qiu, and R.P.S.M. Lobo, ``Optical conductivity of Ba$_{0.6}$K$_{0.4}$Fe$_{2}$As$_{2}$: The effect of in-plane and out-of-plane doping in the superconducting gap,'' Europhys. Lett. {\bf 104}, 47006 (2013). \par 

\bibitem {Bonn88prl} D.A. Bonn, J.D. Garrett, and T. Timusk, ``Far-Infrared Properties of URu$_2$Si$_2$,'' Phys. Rev. Lett. {\bf 61}, 1305--1308 (1988). \par 

\bibitem {Pimenov06prb} A. Pimenov, T. Rudolf, F. Mayr, A. Loidl, A.A. Mukhin, and A.M. Balbashov, ``Coupling of phonons and electromagnons in GdMnO$_3$,'' Phys. Rev. B {\bf 74}, 100403(R) (2006). \par 

\bibitem {Xu08prl} X.S. Xu, M. Angst, T.V. Brinzari, R.P. Hermann, J.L. Musfeldt, A.D. Christianson, D. Mandrus, B.C. Sales, S. McGill, J.-W. Kim, and Z. Islam, ``Charge Order, Dynamics, and Magnetostructural Transition in Multiferroic LuFe$_2$O$_4$,'' Phys. Rev. Lett. {\bf 101}, 227602 (2008). \par 

\bibitem {Miller12prb} K.H. Miller, P.W. Stephens, C. Martin, E. Constable, R.A. Lewis, H. Berger, G.L. Carr, and D.B. Tanner, ``Infrared phonon anomaly and magnetic excitations in single-crystal Cu$_3$Bi(SeO$_3$)$_2$O$_2$Cl,'' \PRB 86, 174104, 2012 \par 

\bibitem {Miller13prb} K.H. Miller, X.S. Xu, H. Berger, V. Craciun, Xiaoxiang Xi, C. Martin, G.L. Carr, and D.B. Tanner, ``Infrared phonon modes in multiferroic single-crystal FeTe$_2$O$_5$Br,'' \PRB 87, 224108, 2013 \par 

\bibitem {LaForge10prb} A.D. LaForge, A. Frenzel, B.C. Pursley, Tao Lin, Xinfei Liu, Jing Shi, and D.N. Basov, ``Optical characterization of Bi$_2$Se$_3$ in a magnetic field: Infrared evidence for magnetoelectric coupling in a topological insulator material,'' Phys. Rev. B {\bf 81}, 125120 (2010). \par 

\bibitem {Akrap12prb} Ana Akrap, Micha\"el Tran, Alberto Ubaldini, J\'er\'emie Teyssier, Enrico Giannini, Dirk van der Marel, Philippe Lerch, and Christopher C. Homes, ``Optical properties of Bi$_2$Te$_2$Se at ambient and high pressures,'' Phys. Rev. B {\bf 86}, 235207 (2012). \par 

\bibitem {Martin13prb} C. Martin, E.D. Mun, H. Berger, V.S. Zapf, and D.B. Tanner, ``Quantum oscillations and optical conductivity in Rashba spin-splitting BiTeI,'' \PRB 87, 041104(R), 2013 \par 

\bibitem {Roessler65bjap} D.M. Roessler, ``Kramers-Kronig analysis of reflection data,'' Brit. J. Appl. Phys. {\bf 16}. 1119--1123 (1965). \par 

\bibitem {Ahrenkiel71josa} R.K. Ahrenkiel, ``Modified Kramers-Kronig Analysis of Optical Spectra,'' J. Opt. Soc. Am. {\bf 61}, 1651-1655 (1971). \par 

\bibitem {Bardwell85jcp} J.A. Bardwell and M.J. Dignam, ``Extensions of the Kramers-Kronig transformation that cover a wide range of practical spectroscopic applications,'' J. Chem. Phys. {\bf 83}, 5468--5478 (1985). \par 

\bibitem {Peiponen91prb} K.-E. Peiponen and E.M. Vartiainen, ``Kramers-Kronig relations in optical data inversion,'' \jourpar Phys. Rev. B, 44, 8301 (1991) \par 

\bibitem {King02josab} Frederick W. King, ``Efficient numerical approach to the evaluation of Kramers-Kronig transforms,'' \josaB 19, 2427--2436, 2002 \par 

\bibitem {Lucarini05} V. Lucarini, J.J. Saarinen, K.-E. Peiponen, and E.M. Vartiainen {\it Kramers-Kronig Relations in Optical Materials Research,\/} (Springer-Verlag, Berlin Heidelberg, 2005). \par 

\bibitem {Kuzmenko05rsi} A. B. Kuzmenko, ``Kramers-Kronig constrained variational analysis of optical spectra,'' Rev. Sci. Instrum. {\bf 76}, 083108 (2005). \par 

\bibitem {Crandles08ao} D. Crandles, F. Eftekhari, R. Faust, G. Rao, M. Reedyk, and F. Razavi, ``Kramers-Kronig-constrained variational dielectric fitting and the reflectance of a thin film on a substrate,'' Appl. Opt. {\bf 47}, 4205--4211 (2008). \par 

\bibitem {Bozovic90prb} Ivan Bozovic, ``Plasmons in cuprate superconductors,'' Phys. Rev. B {\bf 42}, 1969--1984 (1990). \par 

\bibitem {Wu93jp} X. Wu, C.C. Homes, S.E. Burkov, T. Timusk, F.S. Pierce, S.J. Poon, S.L. Cooper, and M.A. Karlow. ``Optical conductivity of the icosahedral quasicrystal Al$_{75.5}$Mn$_{20.5}$Si$_{4}$ and its 1/1 crystalline approximant $\alpha$-Al$_{72.5}$Mn$_{17.4}$Si$_{10.1}$,'' J. Phys.: Condens. Matter {\bf 5}, 5975--5990 (1993). \par 

\bibitem {Kamaras95jap} K. Kamar\'as, K.L. Barth, F. Keilmann, R. Henn, M. Reedyk, C. Thomsen, M. Cardona, J. Kircher, P.L. Richards, and J.L. Stehl\'e, ``The low-temperature infrared optical functions of SrTiO$_3$ determined by reflectance spectroscopy and spectroscopic ellipsometry,'' J. Appl. Phys. {\bf 78}, 1235--1240 (1995). \par 

\bibitem {Puchkov96jp} A.V. Puchkov, D.N. Basov, and T. Timusk, ``The pseudogap state in high- superconductors: an infrared study,'' J. Phys.: Condens. Matter {\bf 8}, 10049--10082 (1996). \par 

\bibitem {Kuzmenko02ssc} A.B. Kuzmenko, F.P. Mena, H.J.A. Molegraaf, D. van der Marel, B. Gorshunov, M. Dressel, I.I. Mazin, J. Kortus, O.V. Dolgov, T. Muranaka, J. Akimitsu, ``Manifestation of multiband optical properties of MgB$_2$,'' Solid State Commun. {\bf 121}, 479--484 (2002). \par 

\bibitem {Carbone06prb} F. Carbone, A. B. Kuzmenko, H. J. A. Molegraaf, E. van Heumen, E. Giannini, and D. van der Marel, ``In-plane optical spectral weight transfer in optimally doped Bi$_2$Sr$_2$Ca$_2$Cu$_3$O$_{10}$ Phys. Rev. B {\bf 74}, 024502 (2006). \par 

\bibitem {Stricker14prl} D. Stricker, J. Mravlje, C. Berthod, R. Fittipaldi, A. Vecchione, A. Georges, and D. van der Marel, ``Optical Response of Sr$_{2}$RuO$_{4}$ Reveals Universal Fermi-Liquid Scaling and Quasiparticles Beyond Landau Theory,'' Phys. Rev. Lett. {\bf 113}, 087404 (2014). \par 

\bibitem {Hwang11prb} J. Hwang, I. Schwendeman, B.C. Ihas, R.J. Clark, M. Cornick, M. Nikolou, A. Argun, J.R. Reynolds, and D.B. Tanner, ``{\it In situ\/} measurements of the optical absorption of dioxythiophene-based conjugated polymers,'' \PRB 83, 195121, 2011 \par 

\bibitem {henke93} B.L. Henke, E.M. Gullikson, and J.C. Davis. X-ray interactions: photoabsorption, scattering, transmission, and reflection at $E=50$--30000 eV, $Z=1$--92, Atomic Data and Nuclear Data Tables {\bf 54}, 181--342 (1993). \par 

\bibitem {henkeweb} \path |http://henke.lbl.gov/optical\_constants/| \par 

\bibitem {Toll56pr} John S. Toll, ``Causality and the Dispersion Relation: Logical Foundations,'' \jourpar Phys. Rev., 104, 1760--1770 (1956) \par 

\bibitem {Kubo72jmp} Ryogo Kubo and Munetake Ichimura, ``Kramers-Kronig Relations and Sum Rules,'' \jourpar J. Math. Phys., 13, 1454 (1972) \par 

\bibitem {Hu89ajp} Ben Yu-Kuang Hu, ``Kramers-Kronig in two lines,'' \jourpar Am. J. Phys., 57, 821--821 (1989) \par 

\bibitem {Milton97prl} G.W. Milton, D.J. Eyre, and J.V. Mantese, ``Finite Frequency Range Kramers-Kronig Relations: Bounds on the Dispersion,'' \jourpar Phys. Rev. Lett., 79, 3062--3065 (1997) \par 

\bibitem {King06jpa} Frederick W King, ``Alternative approach to the derivation of dispersion relations for optical constants,'' \journal J. Phys. A: Math. Gen., 39, 10427, 2006 \par 

\bibitem {Peiponen09rpp} K.-E. Peiponen and J.J. Saarinen, ``Generalized Kramers-Kronig relations in nonlinear optical- and THz-spectroscopy,'' \jourpar Reports on Progress in Physics, 72, 056401 (2009) \par 

\bibitem {Bohren10ejp} Craig F. Bohren, ``What did Kramers and Kronig do and how did they do it?'' \jourpar Eur. J. Phys., 31, 573--577 (2010) \par 

\bibitem {Cauchy} A.L. Cauchy, ``Oeuvres compl\`etes,'' Ser. 1, 4, Paris (1890). \par 

\bibitem {branchcut} This equation appears at first glance to be wrong, but it is not. Recall that the phase shift for non-absorbing materials with $n>1$ is $-\pi $ so that $r$ is negative. The range of $\phi $ is $-\pi \le \phi \le 0$. If I wanted to have a positive phase, I could define $\theta = \phi + \pi $ so that $e^{i\theta } =- e^{i\phi }$. In this case, $$ N = { 1 + \sqrt {\R }e^{i\theta } \over 1- \sqrt {\R }e^{i\theta } } $$ and, of course, $$ \rho e^{i\theta } = {N - 1 \over N + 1}. $$ The range of $\theta $ will be $0 \le \theta \le \pi $. The real part of $\rho e^{i\theta }$ becomes negative when $n < 1$ and $\kappa $ is small, requiring $\theta > \pi /2$. \par 

\bibitem {Andermann65josa} G. Andermann, A. Caron, and David A. Dows, ``Kramers-Kronig Dispersion Analysis of Infrared Reflectance Bands.'' \josa 55, 1210--1212, 1965 \par 

\bibitem {Palik} Edward D. Palik, {\it Handbook of Optical Constants of Solids I and II,} (Academic Press, Orlando, 1985 and 1991). 0 



\bibitem {Bennett65ao} J.M. Bennett, E.J. Ashley, ``Infrared reflectance and emittance of silver and gold evaporated in ultrahigh vacuum,'' Applied Optics, {\bf 4}, 221--224, (1965). \par 

\bibitem {Palik14} B. Dold and R. Mecke, ``\"Uber Gangsmetallen und deren Leigerungen im Infrarot,'' Optik {\bf 22}, 435 (1965). \par 

\bibitem {Vc} There is occasional confusion about the value to use for the volume $V_c$, which is described variously as the ``volume of the unit cell'' or the ``volume of one formula unit.'' Of course any carefully described volume will work, but generally one wants the number of charge carriers per atom or per primitive unit cell. And of course the conventional cell often contains several times more atoms than one would guess from the chemical formula. So the user must decide what he or she wants to determine. It may be the number of carriers per Ag atom in silver metal, the number per buckyball in C$_{60}$, the number per dopant in P-doped Si, or the number per copper atom in YBa$_2$Cu$_3$O$_7$. Once the decision is made, compute the volume in the crystal allocated to the desired quantity. One (almost) failsafe approach is to obtain the density $\rho$ of the crystal and the mass $M$ of the entity one is interested in, such as one Ag atom, 60 C atoms, one silicon atom divided by the dopant concentration, or 1/3 the mass of Y + 2Ba + 3Cu + 7O. Then $V_c = M/\rho $. \par 

\bibitem {Tajima89josa} S. Tajima, H. Ishii, T. Nakahashi, T. Takagi, S. Uchida, M. Seki, S. Suga, Y. Hidaka, M. Suzuki, T. Murakami, K. Oka, and H. Unoki, ``Extensive study of the optical spectra for high-temperature superconducting oxides and their related materials from the infrared to the vacuum-ultraviolet energy region,'' J. Opt. Soc. Am. B {bf 6}, 475--482 (1989). \par 

\bibitem {Doyle68ac} P.A. Doyle and P.S. Turner, ``Relativistic Hartree-Fock X-ray and electron scattering factors,'' Acta Cryst. {\bf A24}, 390--397 (1968). \par 

\bibitem {Cromer68ac} D.T. Cromer and J.B. Mann, ``X-ray scattering factors computed from numerical Hartree-Fock wave functions,'' Acta Cryst. {\bf A24}, 321--324 (1968); Los Alamos Report LA--3816 \par 

\bibitem {Hubbell75} J.H. Hubbell, W.J. Veigele, E.A. Briggs, R.T. Brown, D.T. Cromer, and R.J. Howerton, ``Atomic form factors, incoherent scattering functions, and photon scattering cross sections,'' J. Phys. Chem. Ref. Data {\bf 4}, 471--538 (1975). \par 

\bibitem {Henke81} B.L. Henke, ``Low energy x-ray interactions: photoionization, scattering, specular and Bragg reflection,'' in{ \it Low Energy X-ray Diagnostics\/} edited by D.T. Attwood and B.L. Henke (American Institute of Physics Conf.\ Proc.\ {\bf 75}, New York, 1981), pp.~146--155. \par 

\bibitem {Dreier84} P. Dreier, P. Rabe, W. Malzfeldt, and W. Niemann, ``Anomalous X-ray scattering factors calculated from experimental absorption spectra,'' J. Phys. C: Solid State Phys. {\bf 17}, 3123--3136 (1984). \par 

\bibitem {henke88} B.L. Henke, J.C. Davis, E.M. Gullikson, and R.C.C. Perera, ``A preliminary report on x-ray photoabsorption coefficients and atomic scattering factors for 92 elements in the 10--10,000 eV region,'' Lawrence Berkeley National Laboratory Report LBL-26259 (1988). \par 

\bibitem {Burge93} R.E. Burge, ``The interaction of x-rays,'' in {\it X-ray Science and Technology} edited by A.G. Michette and C.J. Buckley, (Institute of Physics, Bristol, 1993), Chapter 5, pp.\ 160--206. \par 

\bibitem {Chantler00} C.T. Chantler, ``Detailed Tabulation of Atomic Form Factors, Photoelectric Absorption and Scattering Cross Section, and Mass Attenuation Coefficients in the Vicinity of Absorption Edges in the Soft X-Ray $?Z=30$--36, $Z=60$--89, $E=0.1$~keV--10 keV?, Addressing Convergence Issues of Earlier Work,'' J. Phys. Chem. Ref. Data {\bf 29}, 597--1048 (2000). \par 

\bibitem {Kane86pr} P.P. Kane, Lynn Kissel, R.H. Pratt and S.C. Roy, ``Elastic Scattering of Gamma-Rays and X-Rays by Atoms,'' Physics Reports Vol. 140, 75-159 (1986). \par 

\bibitem {Sasaki89} S. Sasaki, ``Anomalous scattering factors (1989) Numerical Tables of Anomalous Scattering Factors Calculated by the Cromer and Liberman Method,'' KEK Report, 88-14, 1-136 \par 

\bibitem {Brennan92} S. Brennan and P.L. Cowan, ``A suite of programs for calculating x-ray absorption, reflection and diffraction performance for a variety of materials at arbitrary wavelengths,'' Rev. Sci. Instrum. {\bf 63}, 850--853 (1992). \par 

\bibitem {Pratt94} R.H. Pratt, Lynn Kissel, and P.M. Bergstrom, Jr., ``New Relativistic S-Matrix Results for Scattering - Beyond the Usual Anomalous Factors/Beyond Impulse Approximation,'' in {\it Resonant Anomalous X-Ray Scattering,\/} edited by G. Materlik, C.J. Sparks and K. Fischer (North-Holland: Amsterdam, 1994). \par 

\bibitem {Kissel95ac} Lynn Kissel, B. Zhou, S.C. Roy, S.K. Sen Gupta, and R.H. Pratt, ``Validity of Form-Factor, Modified-Form-Factor and Anomalous-Scattering-Factor Approximations in Elastic Scattering Calculations,'' Acta Crystallographica {\bf A51}, 271--288 (1995). \par 

\bibitem {Chantler95jpcrd} C.T. Chantler, ``Theoretical Form Factor, Attenuation and Scattering Tabulation for $Z=1$--92 from $E=1$--10 eV to $E=0.4$--1.0 MeV,'' J. Phys. Chem. Ref. Data {\bf 24}, 71--643 (1995). \par 

\bibitem {Cullen97} Dermott E Cullen, John H Hubbel, Lynn Kissel, ``EPDL97 The Evaluated Data Library, 97 Version,'' Lawrence Livermore National Laboratory, UCRL-50400, Vol 6, Rev 5 (Sep 19 1997). \par 

\bibitem {spline} The cubic spline bridge occasionally causes excitement by taking the reflectance in the bridge region over unity or below zero. In these cases it cannot be used. The simple power-law bridges have not caused this trouble. \par 

\bibitem {Brust70prb} David Brust, ``Electronic Structure Effects in the Drude and Interband Absorption of Aluminum,'' Phys. Rev. B {\bf 2}, 818 (1970). \par 

\bibitem {Smith86prb} D.Y. Smith and B. Segall, ``Intraband and interband processes in the infrared spectrum of metallic aluminum,'' Phys. Rev. B {\bf 34}, 5191--5198 (1986). \par 

\bibitem {Turner14} Christopher G. Turner, Paul M. Johns, Evan Thatcher, D.B. Tanner, and Juan C. Nino,``Atomic displacive disorder in Bi$_2$Ti$_2$O$_7$,'' preprint. \par 

\bibitem {Anand14} Naween Anand, Sanal Buvaev, A.F. Hebard, D.B. Tanner, Zhiguo Chen, Zhiqiang Li, Kamal Choudhary, S.B. Sinnott, Genda Gu, and C. Martin, ``Temperature-driven band inversion in Pb$_{0.77}$Sn$_{0.23}$Se: Optical and Hall-effect studies,'' (arXiv:1407.5726 [cond-mat.str-el]. \par 

\bibitem {Martin14} C. Martin, K.H. Miller, S. Buvaev, H. Berger, X.S. Xu, A.F. Hebard, and D.B. Tanner, ``Temperature dependent infrared spectroscopy of the Rashba spin-splitting semiconductor BiTeI,'' arXiv:1209.1656 [cond-mat.mtrl-sci]. \par 


\end{thebibliography}
\end{document}